\documentclass{aa}
\usepackage{graphicx}
\begin{document}
\title{$INTEGRAL$ observation of 3EG J$1736-2908$\thanks{Based on observations with
INTEGRAL, an ESA project with instruments and science data center funded by ESA member
states (especially the PI countries: Denmark, France, Germany, Italy, Switzerland, Spain), 
Czech Republic and Poland, and with the participation of Russia and the USA.}}

\author{G.~Di~Cocco\inst{1}, L.~Foschini\inst{1}, P.~Grandi\inst{1}, G.~Malaguti\inst{1}, A.J. Castro-Tirado\inst{2}, 
S. Chaty\inst{3,4},  A.J. Dean\inst{5}, N.~Gehrels\inst{6}, I.~Grenier\inst{3,4}, W. Hermsen\inst{7}, L. Kuiper\inst{7}, 
N. Lund\inst{8}, F. Mirabel\inst{3}}

\offprints{G. Di Cocco  - \email{dicocco@bo.iasf.cnr.it}}

\institute{Istituto di Astrofisica Spaziale e Fisica Cosmica (IASF) del CNR - Sezione di Bologna, 
Via Gobetti 101, 40129 Bologna Italy
\and
Instituto de Astrof\'{\i}sica de Andaluc\'{\i}a (IAA-CSIC), P.O. Box 03004, E-18080 Granada, Spain 
\and
CEA Saclay, DSM/DAPNIA/SAp (CNRS FRE 2591) F$-91191$ Gif-sur-Yvette Cedex, France
\and
Universit\'e Paris 7 Denis-Diderot 2 place Jussieu 75251 Paris Cedex 05
\and
Department of Physics and Astronomy, University of Southampton, Southampton SO17 1BJ, UK
\and
NASA Goddard Space Flight Center, Laboratory for High Energy Astrophysics, Code 661, Greenbelt, MD
\and
SRON National Institute for Space Research, Sorbonnelaan 2, 3584 CA Utrecht, The Netherlands
\and
Danish Space Research Institute, Juliane Maries Vej 30, 2100 Copenhagen, Denmark }

\date{Received 19 March 2004; accepted 11 June 2004}

\abstract{
The possible identification by INTEGRAL of the EGRET source
3EG~J$1736-2908$ with the active galactic nucleus GRS~$1734-292$ is
discussed.  The latter was discovered in 1990 and later identified with a
Seyfert 1 galaxy.  At the time of the compilation of the 3rd EGRET
Catalog, it was not considered as a possible counterpart of the source
3EG~J$1736-2908$, which remained unidentified.  A detailed multiwavelength
study of the EGRET error circle is presented, by including archival 
radio, soft- and hard-X observations, suggesting that GRS$1734-292$ could be a likely
counterpart of 3EG J$1736-2908$, even though this poses very interesting
questions about the production mechanisms of $\gamma$--rays with energies
greater than 100 MeV.

}

\authorrunning{Di Cocco et al.}

\maketitle

\section{Introduction}

The Third Energetic Gamma--Ray Experiment Telescope (EGRET) Catalogue of
high--energy ($E>100$ MeV) $\gamma$--ray sources (Hartman et al. 1999)  
refers to the observations carried out by the EGRET experiment on-board the
Compton Gamma--Ray Observatory (CGRO) between April 1991 and October 1995.  
It contains 271 point sources, including: 1 solar flare, the Large
Magellanic Cloud, 5 pulsars, the radio galaxy Centaurus A, and 66
high-confidence identifications of blazars.  In addition, 27
lower-confidence blazar identifications are included in the catalog, which
leaves 170 sources unidentified.  Since the publication of the catalog, a
lot of effort has been dedicated to the identification of the 170
unidentified EGRET sources (e.g. Grenier 2003).  While most of the
extragalactic counterparts are blazars, two possible identifications of
EGRET sources with radiogalaxies have been recently proposed (Combi et al.
2003, Mukherjee et al. 2003), in addition to Centaurus A. However, the
multiwavelength counterpart search is hampered by the limited EGRET point
spread function (PSF), which gives error radii of $0.5^{\circ}-1^{\circ}$.

The INTEGRAL satellite (Jensen et al. 2003) was launched on October 17th,
2002. During the first year of scientific operations, more than half of
the INTEGRAL Core Programme (CP) (Winkler et al. 2003) was dedicated to a
deep exposure of the Galactic central radian, and to regular weekly scans
of the Galactic plane.

The IBIS imager (Ubertini et al. 2003), one of the two INTEGRAL main
telescopes, combines for the first time broad-band high-energy coverage
($15$~keV - $10$~MeV) with imaging capability.  IBIS has a large field of
view ($19^{\circ}\times19^{\circ}$ at half response), and a fine angular
resolution of $12'$, sampled in $5'$ pixels in the low energy ($0.015-1$
MeV) layer ISGRI (Lebrun et al. 2003), and in $10'$ pixels in the
high-energy ($0.175-10$ MeV) detector PICsIT (Di Cocco et al. 2003).  
ISGRI reaches a point source location accuracy better than $1'$ for a
$\sim30\sigma$ detection (Gros et al. 2003). These unprecedented
characteristics for a gamma-ray telescope clearly make IBIS one of the
best instruments available to date to search for counterparts of
unidentified EGRET sources.

Here we present the first results of a program set up to search for the
identification of the unidentified EGRET sources located in the Galactic
plane. The first INTEGRAL proposed identification of an unidentified EGRET
source concerns 3EG~J$1736-2908$ and the active galactic nucleus GRS$1734-292$.  
3EG~J$1736-2908$ was originally detected by
EGRET with a $\sim6\sigma$ with a $95$\% confidence circle of
$\theta_{95}=0.62^{\circ}$ in radius, at a position very close to the
Galactic Centre (Hartman et al. 1999). The spectrum showed a photon index
$\Gamma=2.18\pm 0.12$ and a flux for $E>100$~MeV of $(3.3\pm 0.6)\times
10^{-7}$~photons~cm$^{-2}$~s$^{-1}$, averaged over the whole viewing
period.

EGRET observed this source in 23 separate viewing periods (Hartman et al.
1999). The measured fractional variability index, defined as
$\delta=\sigma/\mu$ (Nolan et al. 2003), where $\sigma$ and $\mu$ are the
flux standard deviation and average values respectively, is
$\delta=0.64^{+0.48}_{-0.33}$. This value is larger than the
fractional variability expected for steady sources, $\delta<0.14$, thus
indicating that 3EG~J$1736-2908$ is variable. 
A similar conclusion was independently obtained by McLaughlin et al. (1996) using
a different method to quantify the EGRET flux variability.

It is worth noting that the two established source classes able to generate 
high--energy $\gamma$--rays in the EGRET range, pulsars and active galactic nuclei
(AGN), are known to produce steady and strongly variable fluxes,
respectively. 

\section{INTEGRAL/IBIS observations and data analysis results}
\begin{figure}
\centering
\includegraphics[scale=0.50]{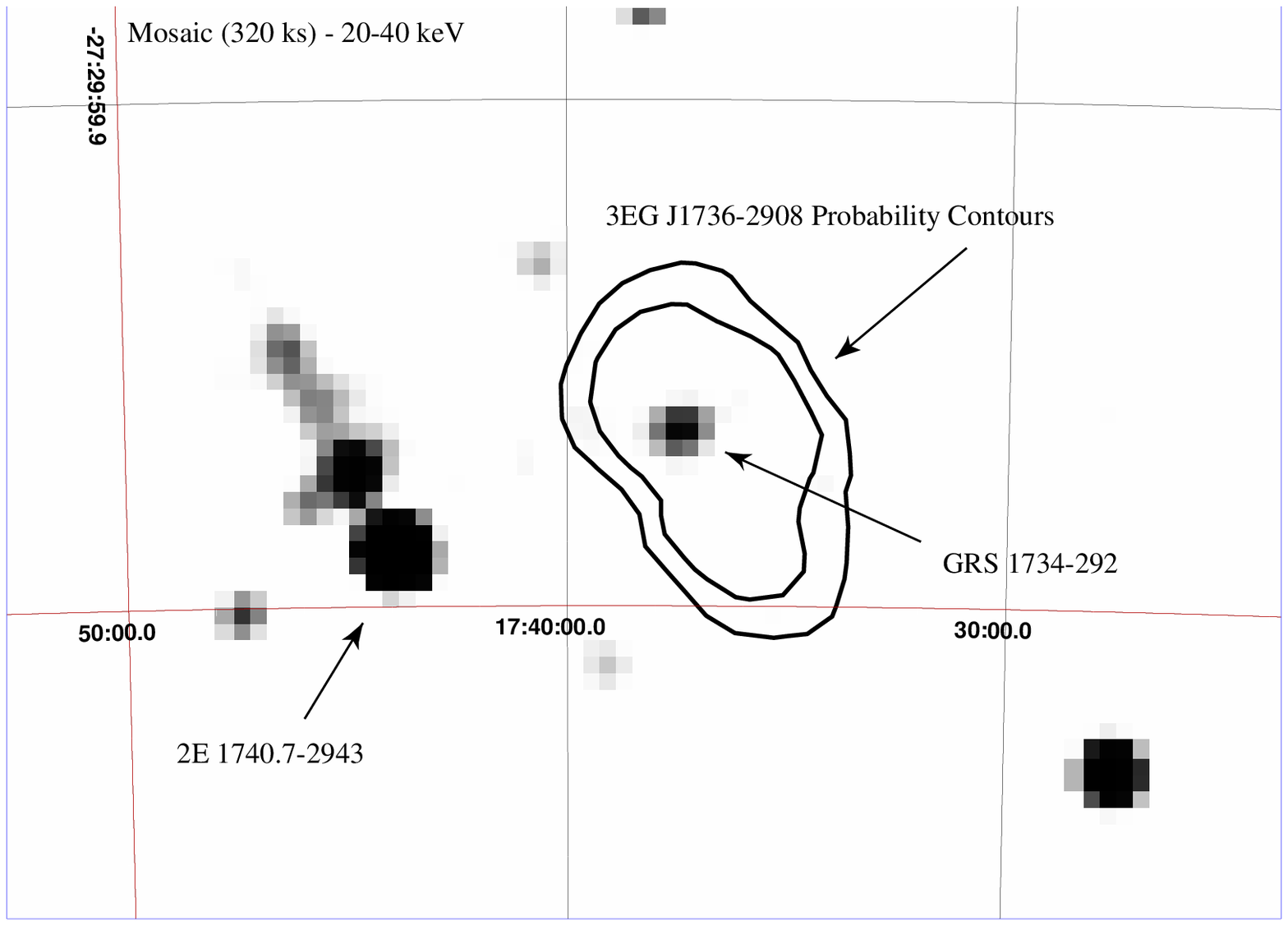}\\
\includegraphics[scale=0.50]{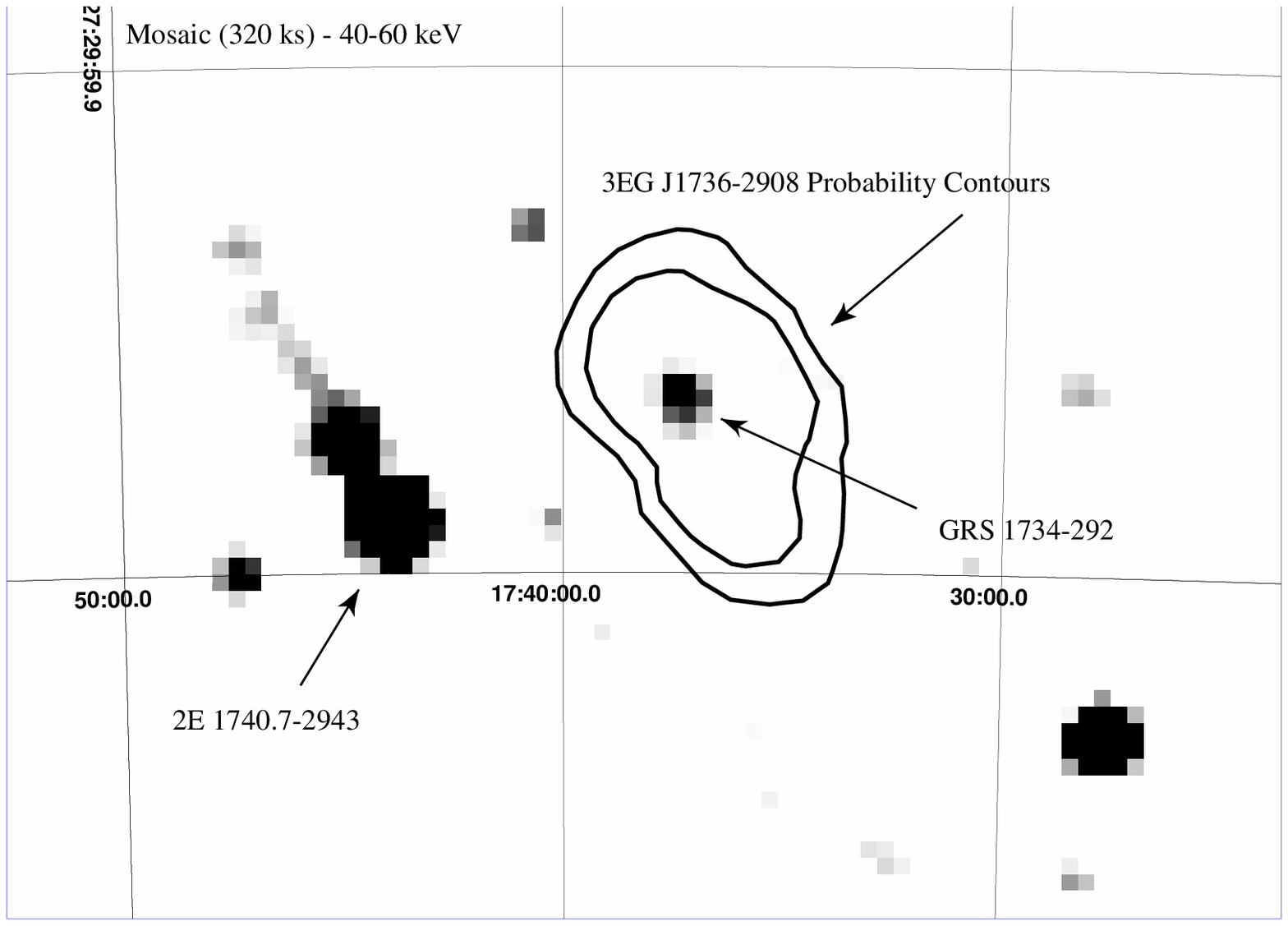}
\vspace*{0.5cm}
\caption{IBIS/ISGRI deconvolved images of the 3EG~$J1736-2908$ sky
region: mosaic of the region with exposure of 320 ks, in the energy bands
20--40 ({\em left}) and 40--60 ({\em right}) keV. The 3EG J$1736-2908$
probability contours at 95\% and 99\% confidence level are shown. The
detected source position inside the probability contours is coincident
with the GRANAT source GRS $1734-292$ (see text)}.
\label{images}
\end{figure}

In order to allow imaging with the SPI instrument (Sch\"onfelder et al. 2003), the
INTEGRAL observing strategy consists of several off-target pointings of
the spacecraft axis following a dithering pattern with steps of 2$^\circ$
(Jensen et al. 2003).  The CP Galactic Plane scans are performed by
INTEGRAL following a sawtooth pattern with $21^{\circ}$ of inclination
with respect to the Galactic plane and with a shift of $27.5^{\circ}$ in
Galactic longitude between two scans (Winkler et al. 2003).  Each scan
consists of an alternated sequence of very short ($\approx$120 s)
slews and staring pointings of $2200$~s (each staring pointing is called
Science Window, ScW).  Sometimes it occurs that in the consolidation
process at the INTEGRAL Science Data Center (ISDC) (Courvoisier et al.
2003), two nearby ScWs are merged, resulting in a longer ScW ($\approx
5000$~s).

The single CP ScWs are sequentially analysed, and all the detected sources
are cross--correlated with the 3rd EGRET Catalogue to look for source
identifications. All the INTEGRAL data analysis described in the present
work has been done with the most recent version of the Offline Standard
Analysis (OSA\footnote{Available at
http://isdc.unige.ch/index.cgi?Soft+download})  v. 3.0 (Goldwurm et al.
2003a). This strategy is useful to search for sources in outburst.

A source inside the error contours of the EGRET
unidentified source 3EG~J$1736-2908$ was found in ScW 57 of revolution 61, 
with a flux of $(2.1\pm0.3)\times10^{-10}$ erg cm$^{-2}$ s$^{-1}$ (or $28\pm
4$~mCrab\footnote{1 mCrab$_{\rm 20-40\;keV} = 7.6\times10^{-12}$ erg
cm$^{-2}$ s$^{-1}$; 1 mCrab$_{\rm 40-60\;keV} = 4.2\times10^{-12}$ erg
cm$^{-2}$ s$^{-1}$, by assuming $F_{\rm Crab}(E)=9.6\cdot
E^{-2.1}$~ph~cm$^{-2}$~s$^{-1}$~keV$^{-1}$.}) in the $20-40$~keV energy
band and signal--to-noise ratio of about $10\sigma$.  To take into account the 
uncertainties in the count rate estimation
of the present version of OSA for ISGRI (see Goldwurm et al.  2003b), we
adopted the procedure outlined by Goldoni et al. (2003) for the analysis
of the off--axis sources. The average fluxes in mCrab have been obtained
using calibration observations of the Crab Nebula at similar off--axis
angles and adding a $5\%$ systematic error.

To strengthen the detection significance and to search for other possible
counterparts within the EGRET error circle, all the ScWs containing the
EGRET region within $10^{\circ}$ from the centre of the field of view
(FOV) were selected and summed together.  The EGRET error circle of
3EG~J$1736-2908$ was observed by INTEGRAL between revolution 53 and 63, in
March-April 2003, for a global exposure of $\approx 320$~ks.  The
resulting images are shown in the upper ($20-40$ keV) and lower ($40-60$
keV) panels of Figure \ref{images}.  The source position ($90\%$
confidence radius = $1.2'$) obtained with this longer exposure,
$\alpha=17:37:27$ and $\delta=-29:08:24$ (J2000), is fully compatible with
the single ScW detection, while the detection significances are now
$17\sigma$, and $8.5\sigma$, in the $20-40$ and $40-60$ keV bands,
respectively, and the associated flux
$(4.9\pm0.4)\times10^{-11}$~erg~cm$^{-2}$~s$^{-1}$ and $(2.1\pm
0.3)\times10^{-11}$~erg~cm$^{-2}$~s$^{-1}$.

The observed differences in the derived flux values could be an indication
for strong flux variability. The source was only detected in ScW 57, while in
all the other ScW it  was below the detection threshold, with a flux upper
limit ($3\sigma$) for a single typical ScW of about
$8\times10^{-11}$erg~cm$^{-2}$~s$^{-1}$.

In this scenario, the positive detection in ScW 57 (elapsed time $\simeq
7000$~s, effective exposure 5000 s) has then to be associated with the
source being in outburst, while in the deep exposure ($\sim320$ ks) the
source becomes visible also in its quiescent state. The robustness of this
result has been confirmed by cross-checking the fractional flux variations
of the three strongest sources in the field. One of these source was
steady, while the other two, although not constant, showed variability
patterns different from the EGRET candidate counterpart.

The centroid position of the IBIS source is $<0.6'$ from the known GRANAT
source GRS $1734-292$, well within the IBIS point source location accuracy
for the observed source strength ($\simeq2'-3'$).  This therefore gives a
first, strong, indication that GRS $1734-292$ is the source detected by
INTEGRAL.

The same source is also clearly visible in the mosaics of the Galactic
Centre region presented by Paizis et al.~(2003), in the field of view of
the black hole candidate H$1743-322$ (Parmar et al.~2003), and in the
IBIS/ISGRI survey of the Galactic Centre region (Revnivtsev et al. 2004).
It is worth noting that in the 2 Ms IBIS/ISGRI mosaic by Revnivtsev et al.
(2004) the source is detected in the $18-60$~keV energy band at 35$\sigma$
level with a flux of $\sim7\times10^{-11}$~erg~cm$^{-2}$~s$^{-1}$,
consistent with the present results. Moreover, it is easy to recognize
that in the 2 Ms IBIS/ISGRI survey there are no other sources in the EGRET
error circle, down to a flux limit of $\simeq1.3\times10^{-11}$~erg~cm$^{-2}$~s$^{-1}$ ($6.5\sigma$).

Finally, the data from the INTEGRAL X-ray instrument JEM-X (energy range: $3-35$ keV; Lund et al. 2003), show
the presence of only one source ($\approx$10 mCrab above 8 keV) within the 95$\%$ confidence contours of 3EG
J$1736-2908$, with a mosaic having effectively a 20 ks exposure time.
The position of this source is compatible with GRS$1734-292$ within the angular resolution of JEM-X at this flux
level ($\approx2'$).

\section{GRS {\boldmath 1734-292}}

GRS $1734-292$ was originally discovered by the coded-mask telescope ART--P (Sunyaev et al. 1990) 
onboard the GRANAT satellite (Sunyaev 1990). Early analysis in the $3-30$~keV energy band showed a power law
spectrum with $\Gamma=2.0\pm0.1$, an absorbing column of
$\simeq6\times10^{22}$~cm$^{-2}$, and the flux at $10$~keV was $(3.2\pm
0.6)\times 10^{-4}$~photons~cm$^{-2}$~s$^{-1}$ (Pavlinsky et al. 1994),
which corresponds to a $20-40$ keV flux of $\sim4\times10^{-11}$ erg
cm$^{-2}$ s$^{-1}$. In 1992, GRS~$1734-292$ experienced an outburst in the
$40-400$~keV energy band, during which a flux variation by a factor
$\sim2$ in a few days was detected with the coded-mask telescope SIGMA (Paul et al. 1991) 
onboard the GRANAT satellite (Churazov et al. 1992). 

Pavlinsky et al. (1994) estimated the $90\%$ confidence level error radius 
of GRS~$1734-292$ as $95'$. This value was later reduced to $3''$ by the 
ROSAT observation of Barret \& Grindlay (1996). The arcsecond level position
allowed to find the optical counterpart.
GRS $1734-292$ was optically classified by Mart\'{\i} et al. (1998) as a
Seyfert 1, with $m_{\rm V}=21.0\pm0.3$, on the basis of the strong and
very broad emission of blended H$\alpha$ and N[II] lines (the FWZI of the
blended line is $\sim2\times10^{4}$~km s$^{-1}$). The redshift was measured
as $z=0.0214$.

In a VLA observation, GRS $1734-292$ was resolved with a clear bipolar
jet-like structure (Mart\'{\i} et al. 1998).  Its radio spectrum could be
well described by a power law $S_\nu\propto\nu^{-\alpha}$ with
$\alpha=0.75\pm0.03$. The total (core plus jets) flux densities at 1.4 and 5 GHz are
$\simeq59$ and $\simeq23$ mJy, respectively.

Sakano et al. (2002) reported the detection of GRS $1734-292$ with ASCA.
The source was observed twice, in 1998 and 1999, with a $0.7-10$ keV flux
of 4.3 and $3.1\times10^{-11}$ erg cm$^{-2}$ s$^{-1}$, respectively. The
spectrum was well fitted by a power law with slope $\Gamma\sim1.4\div1.5$,
absorbed by a column density $N_{\rm H}<2.1\times10^{22}$ cm$^{-2}$. The
value of $N_{\rm H}$ was well constrained by ASCA only in one occasion
giving a value of $1.7^{+0.4}_{-0.3}\times 10^{22}$~cm$^{-2}$. This is
above the galactic hydrogen column density of $\simeq0.8\times10^{22}$
cm$^{-2}$ in the direction of the source measured by Dickey and Lockman
(1990), but consistent with the measurement of the absorption column of
$1.4\times 10^{22}$~cm$^{-2}$ by Schlegel et al. (1998), based on COBE and
IRAS data. Indeed, given the position of the source very close to the
Galactic centre, the measurements of the absorption column have to be
taken with great caution. For example, independent measurements performed
in the optical band with three different methods (Mart\'{\i} et al. 1998)
gave $N_{\rm H}= (1.0\pm 0.2)\times 10^{22}$~cm$^{-2}$.

\section{X-ray and radio cross-correlations and counterparts}

The error circle of 3EG J$1736-2908$ contains no obvious galactic EGRET counterparts.  The apparent association
with the pulsar PSR J$1736-2843$ is almost certainly due to a chance alignment (Kramer et al. 2003) on the basis
of the variability indicator defined by McLaughlin et al. (1996). Moreover, an independent variability indicator
(Nolan et al. 2003) also supports the hypothesis for a non pulsar origin for 3EG J$1736-2908$.
On the other hand, Pulsar Wind Nebulae (PWNe) could be responsible of strong emission in
the EGRET range with a variability consistent with what detected for 3EG J$1736-2908$. 
However none of the PWNe detected in the Galactic Plane falls within the 3EG J$1736-2908$ error circle (Nolan et 
al. 2003, Grenier 2003).

Further possibility of galactic counterparts, which are suggested to be pulsar tracers, can be also excluded.  
No Wolf-Rayet stars, Of stars, SNRs and OB associations are present within the 95$\%$ confidence
contours of 3EG J$1736-2908$ (Romero et al. 1999).

The fractional variability of 3EG J$1736-2908$ is consistent with typical values expected for quasars and BL
Lacs (Nolan et al. 2003).  The principal method of identification of high-latitude EGRET sources relies on
finding positional coincidences with flat spectrum radio loud AGNs.  The NRAO/VLA Sky Survey (NVSS) catalog
(Condon et al. 1998) contains, in the error circle of 3EG J$1736-2908$, 85 radio sources. All the sources are
radio-weak, with typical fluxes around $20-50$ mJy at 1.4 GHz. Even in the assumption of a flat radio spectrum,
this implies weak 5 GHz flux densities $<200$ mJy, i.e. below the lowest fluxes measured for identified EGRET
blazars (Wallace et al. 2002). Torres et al. (2001) have explored the error circles of all low galactic latitude
unidentified EGRET sources, listing all significant point-like radio-sources having flux densities above 30 mJy
at least at one wavelength. The only two strong radio emitters they found are: PNESO 359.3+01.4 and
OH359.140+01.137.  The first is a planetary nebula, while the latter is a molecular cloud. Both sources can be
excluded as responsible for the gamma-ray emission on the basis of the observed EGRET variability (Nolan et al.
2003).

We have also cross-correlated all the ROSAT X-ray satellite catalogs
available in the HEASARC archive with the 95\% confidence EGRET error
radius of 3EG J$1736-2908$. This resulted in $16$ sources, only one of
which has a radio counterpart in the NRAO/VLA Sky Survey (NVSS) catalog
(Condon et al. 1998):  the GRANAT source GRS $1734-292$.

The Galactic Centre region was also scanned by the ASCA X-ray satellite
(see Section 3, Sakano et al.  2002). We have searched the 3EG
J$1736-2908$ error circle and found five ASCA sources, only two of which
have a radio counterpart in the NVSS catalog: again GRS$1734-292$, and
AXJ$1734.5-2915$. The latter is very weak both in radio and X-rays, with a
1.4 GHz flux density of 3.4 mJy, while the ASCA X-ray flux ($F_{\rm
0.7-10\;keV}=3\times10^{-13}$ erg cm$^{-2}$ s$^{-1}$) is a factor
$\simeq$100 less than what was measured for GRS $1734-292$.

We thus conclude that the EGRET error circle does not contain any obvious
blazar-like candidate. The only source which shows a broad band spectrum
from radio to soft (ROSAT) and hard (ASCA, GRANAT, INTEGRAL)  X-rays, is
GRS $1734-292$.

\begin{figure*}[!ht]
\centering
\includegraphics[scale=0.6, angle=90]{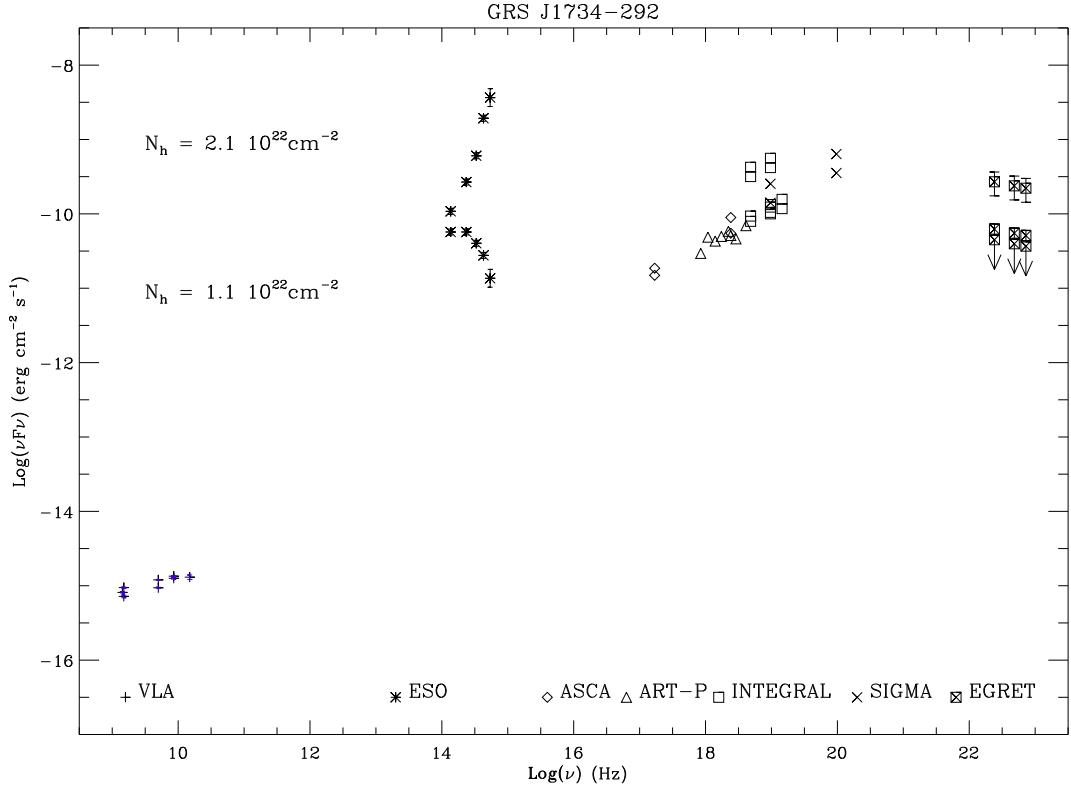}
\vspace*{0.5cm}
\caption{Spectral Energy Distribution (SED) of GRS~$1734-292$.
We show here all published data: VLA and ESO (Mart\'{\i} et al. 1998), ASCA (Sakano et al. 2002), 
ART-P (Pavlinsky et al. 1994), INTEGRAL (present paper), SIGMA (Churazov et al. 1992) and EGRET (Hartman et al. 1999).
Infrared and optical data from ESO are corrected with two different absorptions, 
respectively $1.1\times 10^{22}$ and $2.1\times 10^{22}$~cm$^{-2}$.
The data include minimum, mean and maximum observed values, in order to show the interval of 
variation of GRS~$1734-292$ from non--detection to flares, at different epochs.
If not visible, error bars are smaller than size of symbols.}
\label{sed}
\end{figure*}

\section{Discussion}
During its Galactic Plane scans, the IBIS imager and the JEM-X monitor
on-board INTEGRAL detected only one source within the 3EG J$1736-2908$ error 
circle. In IBIS the source was first detected, between 20 and 40 keV in a 
snap-shot ($\sim$5000 s) observation in one ScW only. A longer integration of 
$320$~ks allowed the detection of the source at much higher significance and 
also in the $40-60$ keV band. ScW level monitoring suggests a possible short-time 
(few hours) variability.

The INTEGRAL results together with the measured EGRET variability (Nolan
et al. 2003) seem to indicate an identification of 3EG J$1736-2908$ with
GRS $1734-292$.  Moreover, the multiwavelength search in the probability
contours of 3EG J$1736-2908$ did not reveal the presence of possible
blazar-like radio-loud objects with spectral characteristics similar to
the already identified EGRET sources. This scenario would leave 
GRS~$1734-292$ as the most likely AGN to be associated with the EGRET source.
Sazonov et al. (2004) have recently calculated the probability to
find by chance an EGRET source consistent with the position of GRS$1734-292$
to be only $3\%$. 

The non-classical blazar nature of GRS $1734-292$ is not necessarily a
problem. In fact, the radiogalaxy NGC 6251, showing a jet forming a large
angle with the line of sight, has been proposed for an association with
the EGRET source 3EG J$1621+8203$ (Mukherjee et al. 2002). However, this
object is a strong radio emitter, with a core 1.4 GHz flux density $>$ 800
mJy, and its spectral energy distribution is well fitted, from radio to
gamma-rays, by a double peak synchrotron-self-Compton (SSC)  model typical
of jet-dominated AGNs (Guainazzi et al. 2003, Chiaberge et al. 2003). On
the other hand, an association of another EGRET source, 3EG J$1735-1500$
with a radio-faint galaxy has recently been proposed (Combi et al 2003).

The observed radio faintness of GRS $1734-292$, $F_{1.4\;GHz}\simeq50$
mJy, combined with the steepness of the radio spectrum, makes it difficult
to explain the EGRET emission in terms of the inverse Compton upscattered
component from a jet.  In addition, the derived spectral index connecting
the radio and optical bands is $\alpha_{\rm RO}<0.16$ (Sambruna et al.
1996). This value characterizes GRS$1734-292$ as a radio-quiet source
(radio-loud objects have by definition $\alpha_{\rm RO}>0.2$). Note, for
example, for NGC 6251, $\alpha_{\rm RO}=0.7$.  However, the $\alpha_{\rm
RO}$ value strongly depends on the assumed line-of-sight optical
absorption, and since GRS 1734-292 lies close to the Galactic Centre, the
uncertainty in its associated $A_{\rm V}$ value is high. In fact, in the
most conservative scenario, i.e. assuming a value for $A_{\rm V}$ as low
as 4, corresponding to $N_{\rm H}=0.8\times10^{22}$ cm$^{-2}$ (Dickey and
Lockman 1990), the resulting spectral index is $\alpha_{\rm RO}=0.32$,
thus satisfying the condition for radio loudness.

On the other hand, the Spectral Energy Distribution (SED) of GRS $1734-292$ shown in Fig.~\ref{sed} is intriguing. 
We point out that radio fluxes shown in the SED include the contribution of the whole radio source, i.e. both 
core and jets. Since the jet emission is mainly of synchrotron nature, the radio slope would be steeper by taking 
into account only the core of the source, and the radio spectrum would then be more naturally aligned towards the 
infrared data, pointing towards a synchrotron origin. It appears from the SED that ASCA, ART-P, INTEGRAL and
SIGMA spectra are fully compatible. Multi-wavelength data, from optical to gamma-ray, resemble a typical 
two-peak blazar shape. The first peak will be in the near-infrared/optical domain, however its precise location 
is hampered by the uncertainty in the column density, as is visible from the two infrared/optical spectra derived 
with different estimates of column density. The second peak will be between SIGMA and EGRET data. Simultaneous 
multi-wavelength data would help in constraining the accurate shape of GRS 1734-292's SED.


In conclusion, the INTEGRAL scans of the Galactic Centre regions indicate that GRS $1734-292$ is a possible
counterpart candidate of the unidentified EGRET source 3EG J$1736-2908$.  GRS $1734-292$, optically classified
as Seyfert 1 and weak radio emitter, is difficult to reconcile with a picture, in which the radio to
$\gamma-$ray emission is completely explained in term of strongly beamed non-thermal radiation.  In the case
that future deeper and higher angular resolution observations in the EGRET band (GLAST, AGILE) will confirm the
present scenario, new possible models for $\gamma$-ray emission from AGN will have to be explored.

\begin{acknowledgements} The italian participation to the INTEGRAL/IBIS project is financed by the
Italian Space Agency (ASI).  This work has made use of public data obtained from the High Energy
Astrophysics Science Archive Research Centre (HEASARC), provided by NASA Goddard Space Flight Centre.

\end{acknowledgements}

\end{document}